\documentclass[journal,draftclsnofoot,onecolumn]{IEEEtran}

\usepackage{amssymb}

\usepackage{amsmath}

\usepackage{cite} 

\usepackage{mathrsfs}

\usepackage{graphicx}

\usepackage[square, comma, sort&compress, numbers]{natbib}

\begin{document}

\newtheorem{theorem}{Theorem}
\newtheorem{proposition}{Proposition}
\newtheorem{lemma}{Lemma}
\newtheorem{definition}{Definition}
\newtheorem{corollary}{Corollary}
\newtheorem{remark}{Remark}
\newtheorem{example}{Example}

\title{Necessities and sufficiencies of a class of permutation polynomials over finite fields
 }

\author{Xiaogang~Liu  
\thanks{
X. Liu is with
College of Computer Science and Technology,
Nanjing Tech University, 
Nanjing City,
Jiangsu Province,
PR China
211800
     e-mail:liuxg0201@163.com. }
}

\maketitle

\begin{abstract}
For the finite field $\mathbb{F}_{2^{3m}}$, permutation polynomials of the form $(x^{2^m}+x+\delta)^{s}+cx$ are studied. Necessary and sufficient conditions are given for the polynomials to be   permutation polynomials. For this, the structures and properties of the field elements are analyzed.

\end{abstract}

 \begin{IEEEkeywords}
Finite field; \ Permutation polynomial; \ Trace function
 \end{IEEEkeywords}

\section{Introduction}\label{secI}
 
Let $\mathbb{F}_q$ be a finite field with $q$ elements, where $q$ is a power of prime $p$. A polynomial $f(x)\in \mathbb{F}_q$ is called a permutation polynomial (PP) of $\mathbb{F}_q$, if the induecd mapping $f:c\rightarrow f(c)$ is a bijection of $\mathbb{F}_q$. Earlier work on PPs  can be traced back to  \cite{C1}. Now they are an interesting subject of research recently. and they have important applications in coding theory, cryptography and combinatorial designs \cite{ FF01,M1,V06}. For more details and recent advances and contributions, we refer the reader to  \cite{D01,D02,H1,H2,L1,L2,PL,ZZH1}. 

A  monomial $x^n$ permutes $\mathbb{F}_q$ if and only if $\textup{gcd}(n,q-1)=1$, they are the simplest type of permutation polynomials. For binomials and trinomials, it is not so easy to decide whether they are PPs.  In \cite{ZYY}, D. Zheng {\sl et al.} studied permutation polynomials of the form $f(x)=g(x^{q^k}-x+\delta)+cx$, and they proved that $f(x)$ is a permutation polynomial of $\mathbb{F}_{q^m}$ if and only if $h(x)=g(x)^{q^k}-g(x)+cx$ permutes $\mathbb{F}_{q^m}$ where $c$ belongs to some subfield of $\mathbb{F}_{q^m}$.  In their results $f(x)$ is connected with $\delta$ which can be any element of $\mathbb{F}_{q^m}$, but $h(x)$ is not connected with $\delta$.  In \cite{LWLZ}, L. Li {\sl et at.} considered polynomials of the form $f(x)=\sum\limits_{j=1}^t(x^{p^m}-x+\delta)^{s_j}+x$, and  L. Li {\sl et at.} related their permutation properties over $\mathbb{F}_{p^n}$ with permutation properties over a subset of $\mathbb{F}_{p^n}$ of polynomials of the form $\sum\limits_{j=1}^t((x+\delta)^{p^ms_j}-(x+\delta)^{s_j})+x$. In particular, \cite{LWLZ} showed that $(x^{2^m}+x+\delta)^{s}+cx$ is a PP over $\mathbb{F}_{2^{3m}}$ when $s=2^{2m-1}+2^{m-1}$ and $c=1$, where $\textup{gcd}(m-1,3m)=1$. In this paper, we continue their work with   general $c\in \mathbb{F}_{2^{m}}$ for odd integer $m$. Necessary and sufficient conditions are given. Especially for   $c=1$, as in \cite{LWLZ}, we showed that it is a PP when $m=3k+2$, but it is not a PP when $m=3k+1$. 

First, we present some notations and a lemma which might be useful. 
For two positive integers $m,n$ with $m|n$, a subset $\mathcal{J}$ of $\mathbb{F}_{2^n}$ is defined by
\begin{equation*}\label{e01}
\mathcal{J}=\{\gamma^{2^m}+\gamma |\gamma\in \mathbb{F}_{2^n}\}.
\end{equation*}
 In the following, we consider the case of $n=3m$.  The $d$th root of unity is defined by
\[
\mu_d=\{x\in \mathbb{F}_{2^n}: x^d=1 \}.
\] 
 
\begin{lemma}\cite{BRS}\label{l01}
For a positive integer $m$, let $\alpha, \beta \in \mathbb{F}_{2^m}^* $. The cubic equation $x^3+\alpha x+\beta=0$ has
\begin{enumerate}
\renewcommand{\labelenumi}{$($\mbox{\roman{enumi}}$)$}
\item
  exactly one root in $\mathbb{F}_{2^m}$ if and only if $\textup{Tr}_{2^m/2}(1+\alpha^3\beta^{-2})=1$;
\item
three distinct solutions in $\mathbb{F}_{2^m}$ if and only if $p_m({{\beta}\over {\alpha^{3\over 2}}})=0$, where the polynomial $p_m(x)$ is recursively defined by $p_1(x)=p_2(x)=x, p_k(x)=p_{k-1}+x^{2^{k-3}}p_{k-2}$ for $k\geq 3$;
\item
no solutions in $\mathbb{F}_{2^m}$, otherwise.
\end{enumerate}
\end{lemma}

\section{Main results}\label{SecII}
In this section, we present the main results of our study in Theorem \ref{p01}. For this, field properties and structures of elements of finite fields will be useful for our study. For more properties of finite fields, see \cite{LN1}.
\begin{theorem}\label{p01}
Let $3\nmid m$ be a positive odd integer, and  $\delta \in \mathbb{F}_{2^{3m}}, c\in \mathbb{F}_{2^{m}}^*$. The polynomial
\[
f(x)=(x^{2^m}+x+\delta)^{2^{2m-1}+2^{m-1}}+cx
\]
\begin{enumerate}
\renewcommand{\labelenumi}{$($\mbox{\roman{enumi}}$)$}
\item
  is a PP of $\mathbb{F}_{2^{3m}}$ if and only  $p_m({{c^2+1}\over {  1+c+c^2}})\not=0$ when $m=3k+1$;
\item
 is a PP of $\mathbb{F}_{2^{3m}}$ if and only if  $p_m({{1}\over {  1+c+c^2}}) \not=0$ when $m=3k+2$.
\end{enumerate}
\end{theorem}

\begin{IEEEproof}
In the following, we prove the $m=3k+1$ case, and the $m=3k+2$ will be mentioned at the end.

 Sufficiencies:

 As in the proof of \cite[Proposition 4]{LWLZ}, we need to show that for any $d_0 \in \mathcal{J}$,
\begin{equation*}\label{e02}
x^{2^{2m+1}}+\textup{Tr}_m^{3m}(\delta)x^{2^{2m}}+c^2x^2+\delta^{2^{2m}}(\delta^{2^m}+\delta)=d_0^2
\end{equation*}
has at most one solution in $\mathcal{J}$. Set $\delta_0=\textup{Tr}_m^{3m}(\delta)\in \mathbb{F}_{2^m}$. If $x_1\not= x_2$ are both solutions  of the above equation, then $x=x_1+x_2$ satisfy
\begin{equation}\label{e03}
x^{2^{2m+1}}+\delta_0 x^{2^{2m}}+c^2x^2=0.
\end{equation}
Note that if $x_1=z_1^{2^m}+z_1, x_2=z_2^{2^m}+z_2 \in \mathcal{J}$, then
\begin{equation}\label{e032}
x=(z_1+z_2)^{2^m}+(z_1+z_2) \in \mathcal{J}.
\end{equation}
If $\delta_0=0$,  equation (\ref{e03}) becomes
\begin{equation*}\label{e04}
x^{2^{2m+1}}=c^2x^2. 
\end{equation*}
That is 
\begin{equation}\label{e05}
x^{2^{2m }}=c x. 
\end{equation}
Taking the $2^m$th power for the above equation 
\begin{equation}\label{e06}
x =c x^{2^m}. 
\end{equation}
Substiting the above equation into  (\ref{e05}) 
\begin{equation*}\label{e07}
x^{2^{2m }}=c^2x^{2^m}. 
\end{equation*}
Taking the $2^m$th power for  the above equation 
\begin{equation*}\label{e08}
x =c^2 x^{2^{2m}}. 
\end{equation*}
Substituting the above equation into  (\ref{e05}) 
\begin{equation*}\label{e09}
 c^3 =1,
\end{equation*}
for $x $ is nonzero. Since $m$ is odd, $\textup{gcd}(2^m-1,3)=1$, we get
\begin{equation*}\label{e10}
 c =1.
\end{equation*}
And equation (\ref{e06}) beomces
\begin{equation}\label{e11}
x = x^{2^m}. 
\end{equation}

Since $x\in \mathcal{J}$ by equation (\ref{e032}), we can assume that
\begin{equation}\label{e12}
x=z^{2^m}+z
\end{equation}
for some $z\in \mathbb{F}_{2^{3m}}$. Substituting into equation (\ref{e11}) 
\begin{equation*}\label{e13}
 z^{2^m}+z=z^{2^{2m}}+z^{2^m}.
\end{equation*}
That is 
\begin{equation*}\label{e14}
  z=z^{2^{2m}}. 
\end{equation*}
Taking the $2^m$th power of the above equation, we have
\begin{equation*}\label{e15}
  z=z^{2^{ m}}. 
\end{equation*}
Thus $z\in \mathbb{F}_{2^{m}}$, and from equation (\ref{e12}) 
\begin{equation*}\label{e16}
x=0,
\end{equation*}
contradiction.

In the following, we consider the case $\delta_0\not=0$. Let $x=\delta_0 y$, then $\delta_0\in \mathbb{F}_{2^m}^*$, and equation (\ref{e03}) becomes
\begin{equation*}\label{e17}
\delta_0^2y^{2^{2m+1}}+\delta_0^2 y^{2^{2m}}+c^2\delta_0^2y^2=0.
\end{equation*}
Dividing the above equation by $\delta_0^2$, we have 
\begin{equation}\label{e20}
 y^{2^{2m+1}}+ y^{2^{2m}}+c^2 y^2=0.
\end{equation}
If $x=z^{2^m}+z$, then
\begin{equation}\label{e18}
y={x\over \delta_0}={{z^{2^m}+z}\over \delta_0}=({z\over \delta_0})^{2^m}+{z\over \delta_0}.
\end{equation}
Thus $y\in \mathcal{J}$. 

Since $m$ is odd, $\textup{gcd}(2^m-1,2^{2m}+2^m+1)=1$. Then $2^{3m}-1=(2^m-1)(2^{2m}+2^m+1)$, and every
 $y\in \mathbb{F}_{2^{3m}}^*$ can be written uniquely in the following form
\begin{equation}\label{e19}
y=uv
\end{equation}
where $u\in \mathbb{F}_{2^m}^*$, and $v\in \mu_{2^{2m}+2^m+1}$. As in equation (\ref{e18}), we can find that $v\in \mathcal{J}$. 

For equation (\ref{e20}), take the $2^m$th power 
\begin{equation}\label{e22}
 y^2+ y+  c^2 y^{2^{m+1}}=0.
\end{equation}
Substituting equation (\ref{e19}) into the above equation 
\begin{equation*}\label{e21}
 u^2v^2+ uv +c^2 u^2v^{2^{m+1}}=0.
\end{equation*}
Dividing $c^2u^2$ on both sides of the above equation,  we get
\begin{equation*}\label{e21}
 v^{2^{m+1}}={v^2\over c^2}+ {1\over c^2}{v\over u}.  
\end{equation*}
Let $d^2={1\over c^2}, w={1\over u}$, then $d,w\in \mathbb{F}_{2^m}^*$. And the above quation becomes
\begin{equation}\label{e23}
 v^{2^{m+1}}={d^2v^2}+ {d^2}{wv}.  
\end{equation}
Taking the $1\over 2$th power for the above equation 
\begin{equation}\label{e24}
 v^{2^{m }}={dv}+ {d}{w^{1\over 2}v^{1\over 2} }.  
\end{equation}
Taking the $2^m$th power on both sides of the above equation 
\begin{equation}\label{e25}
\begin{array}{ll}
 v^{2^{2m }}&={dv^{2^m}}+ {d}{w^{1\over 2}(v^{2^m})^{1\over 2} } \\
              &={d({dv}+ {d}{w^{1\over 2}v^{1\over 2} }  )}+ {d}{w^{1\over 2}({dv}+ {d}{w^{1\over 2}v^{1\over 2} }  )^{1\over 2} } \\
&=d^2v+(d^2w^{1\over 2}+d^{3\over 2}w^{1\over 2})v^{1\over 2}+d^{3\over 2}w^{3\over 4}v^{1\over 4}.
 \end{array}
\end{equation}

Since $v\in \mathcal{J}$, we can find that
\begin{equation}\label{e26}
v+v^{2^m}+v^{2^{2m}}=0.
\end{equation}
Substituting equations (\ref{e24}) and (\ref{e25}) into the above equation 
\begin{equation*}\label{e27}
v+({dv}+ {d}{w^{1\over 2}v^{1\over 2} }  )+(d^2v+(d^2w^{1\over 2}+d^{3\over 2}w^{1\over 2})v^{1\over 2}+d^{3\over 2}w^{3\over 4}v^{1\over 4})=0.
\end{equation*}
That is
\begin{equation*}\label{e28}
(1+d+d^2)v+  w^{1\over 2}d(1+d^{1\over 2}+d)v^{1\over 2}    + d^{3\over 2}w^{3\over 4}v^{1\over 4}=0.
\end{equation*}
Note that $1+d+d^2\not=0$ for $m$ is odd. Dividing $1+d+d^2$ on both sides of the above equation, we have
\begin{equation*}\label{e29}
 v+  {w^{1\over 2}d \over {1+d^{1\over 2}+d}}v^{1\over 2}    + {{d^{3\over 2}w^{3\over 4}}\over {(1+d^{1\over 2}+d)^2}}v^{1\over 4}=0.
\end{equation*}
Taking the $4$th power of the above equation 
\begin{equation*}\label{e30}
 v^4+  {w^{  2}d^4 \over {1+d^{  2}+d^4}}v^{ 2}    + {{d^{6 }w^{3 }}\over {(1+d^{  2}+d^4)^2}}v =0.
\end{equation*}
For the above equation, let $d_1={w^{  2}d^4 \over {1+d^{  2}+d^4}}, d_2={{d^{6 }w^{3 }}\over {(1+d^{  2}+d^4)^2}}$, and dividing $v$, we have 
\begin{equation}\label{e311}
 v^3=  d_1v    +d_2. 
\end{equation}

We can find that the following polynomial
\begin{equation*}\label{e32}
t^3+t+1 
\end{equation*}
is irreducible over $\mathbb{F}_2$.  Since $\textup{gcd}(3,m)=1$, it is also irreducible over $\mathbb{F}_{2^m}$. Thus every element $v\in \mathbb{F}_{2^{3m}}^*$ can be written in the following form
\begin{equation}\label{e33}
v=a_0+a_1t+a_2t^2
\end{equation}
with $a_0,a_1,a_2\in \mathbb{F}_{2^m}$, and 
\begin{equation}\label{e34}
t^3=t+1.
\end{equation}
Multiply $t$ on both sides  
\begin{equation*}\label{e35}
t^{2^2}=t^4=t^2+t.
\end{equation*}
Squaring both sides 
\begin{equation*}\label{e36}
t^{2^3} =t^4+t^2=t.
\end{equation*}
Sqaure again, we have
\begin{equation*}\label{e37}
t^{2^4} =t^2.
\end{equation*}
In general, 
\begin{equation}\label{e38}
\left\{
\begin{array}{ll}
t^{2^{3k^{\prime}}}& =t,\\
   t^{2^{3k^{\prime}+1}}& =t^2,\\
  t^{2^{3k^{\prime}+2}}& =t^2+t
 \end{array}
\right.
\end{equation}
for a positive integer $k^{\prime}$. Since $m=3k+1$, we have
\begin{equation}\label{e39}
\left\{
\begin{array}{ll}
t^{2^{m}}& =t^2,\\
   t^{2^{2m}}& =t^2+t.
 \end{array}
\right.
\end{equation}

Substituting  equation (\ref{e39}) into equaiton (\ref{e33}), we get
\begin{equation}\label{e40}
\begin{array}{ll}
v^{2^{m}}& =a_0+a_1t^{2^{m}}+a_2t^{2^{m+1}}\\
   &=a_0+a_1t^2+a_2(t^2+t)\\
&=a_0+a_2t+(a_1+a_2)t^2,
 \end{array}
\end{equation}
and 
\begin{equation}\label{e41}
\begin{array}{ll}
v^{2^{2m}}& =a_0+a_1t^{2^{2m}}+a_2t^{2^{2m+1}}\\
   & =a_0+a_1(t^2+t)+a_2t\\
&=a_0+(a_1+a_2)t+a_1t^2.
 \end{array}
\end{equation}
Substituting the above two equalities into equation (\ref{e26}) 
\begin{equation*}\label{e42}
\begin{array}{ll}
v+v^{2^m}+v^{2^{2m}}&=(a_0+a_1t+a_2t^2)+(a_0+a_2t+(a_1+a_2)t^2)+(a_0+(a_1+a_2)t+a_1t^2)\\
   & =a_0 =0.\\
 \end{array}
\end{equation*}
That is $v$ can be written as 
\begin{equation}\label{e43}
v= a_1t+a_2t^2
\end{equation}
with $a_1, a_2$ not both zeroes.

From our definition $v\in \mu_{2^{2m}+2^m+1}$, that is 
\begin{equation*}\label{e44}
\begin{array}{ll}
v^{1+{2^m}+{2^{2m}}}&=vv^{2^m}v^{2^{2m}}=1.
 \end{array}
\end{equation*}
Substituting equations (\ref{e40}), (\ref{e41}) and (\ref{e43}) into the above equation, we have
\begin{equation*}\label{e45}
\begin{array}{ll}
(a_2 +(a_1+a_2)t)((a_1+a_2)+a_1t)(a_1+a_2t)t^3=1.
 \end{array}
\end{equation*}
Combining with equation (\ref{e34}) 
\begin{equation*}\label{e46}
\begin{array}{ll}
((a_1+a_2)a_1t^2+(a_1a_2+a_1^2+a_2^2)t+a_2(a_1+a_2))(a_2t^2+(a_1+a_2)t+a_1)=1.
 \end{array}
\end{equation*}
That is
\begin{equation*}\label{e47}
\begin{array}{ll}
&(a_1+a_2)a_1a_2t^4+(a_2a_1^2+a_2^3+a_1^3)t^3\\
 +&(a_2a_1^2+a_1a_2^2)t^2+(a_2^3+a_1^3+a_1a_2^2)t\\
+&(a_1+a_2)a_1a_2+1=0
 \end{array}
\end{equation*}
Substituting equations (\ref{e34}) and (\ref{e38}) into the above equation 
\begin{equation}\label{e48}
a_1^3+a_2^3+a_1a_2^2+1=0.
\end{equation}

For equation (\ref{e311}), consider the cubic power of (\ref{e43})
\begin{equation*}\label{e49}
 v^3= (a_1+a_2t)^3t^3. 
\end{equation*}
Combining with equation (\ref{e34}) 
\begin{equation*}\label{e50}
\begin{array}{ll}
 v^3&= (a_1^2+a_2^2t^2)(a_1+a_2t)(t+1)\\
&=(a_1^2+a_2^2t^2)(a_2t^2+(a_1+a_2)t+a_1)\\
&=a_2^3t^4+a_2^2(a_1+a_2)t^3+(a_1a_2^2+a_2a_1^2)t^2+(a_1+a_2)a_1^2t+a_1^3.
 \end{array}
\end{equation*}
Substituting equation (\ref{e38}), we have
\begin{equation*}\label{e51}
\begin{array}{ll}
 v^3&=(a_2^3+a_1a_2^2+a_2a_1^2)t^2+(a_2^3+(a_1+a_2)a_2^2+(a_1+a_2)a_1^2)t+(a_1+a_2)a_2^2+a_1^3\\
    &=(a_2^3+a_1a_2^2+a_2a_1^2)t^2+(a_1^3+ a_1a_2^2+a_2a_1^2)t+a_2^3+a_1^3+a_1a_2^2.
 \end{array}
\end{equation*}
And,  
\begin{equation*}\label{e52}
\begin{array}{ll}
 d_1v+d_2&=d_1(a_1t+a_2t^2)+d_2\\
    &= d_1a_2t^2+d_1a_1t+d_2.
 \end{array}
\end{equation*}
Comparing equation (\ref{e311}) with the above two equations, we get
\begin{equation*}\label{e53}
\left\{
\begin{array}{ll}
 a_2^3+a_1a_2^2+a_2a_1^2 &= d_1a_2, \\
    a_1^3+ a_1a_2^2+a_2a_1^2& =d_1a_1,\\
a_2^3+a_1^3+a_1a_2^2&=d_2.
 \end{array}
\right.
\end{equation*}

The first two of the above equalites, and the fact that $a_1$ and $a_2$ can not be zeroes simultaneously imply that
\begin{equation}\label{e54}
d_1=a_1^2+a_2^2+a_1a_2.
\end{equation}
The third one and equation (\ref{e48}) imply that
\begin{equation*}\label{e55}
d_2=1.
\end{equation*}

From the definitions of $d_1$ and $d_2$, we have
\begin{equation*}\label{e56}
{{d^{6 }w^{3 }}\over {(1+d^{  2}+d^4)^2}}=1.
\end{equation*}
Since $\textup{gcd}(2^m-1,3)=1$, we can take the $1\over 3$ power of the above equation, and get
\begin{equation*}\label{e57}
w= {{(1+d^{  2}+d^4)^{2\over 3}}\over {d^2}}.
\end{equation*}
Thus
\begin{equation*}\label{e58}
d_1={w^{  2}d^4 \over {1+d^{  2}+d^4}}={(1+d^{  2}+d^4)^{1\over 3}}.
\end{equation*}

For $c=1$, that is $d=1$, using  relation (\ref{e38}), we can find that equation  (\ref{e22}) is satisfied for 
\[
y=t+t^2,
\]
that is $f(x)$ is not a PP. In the following, we assume that $d\not=0,1.$


For equations (\ref{e22}) and (\ref{e23}), we find that
\begin{equation*}\label{eb1}
 (a_1t+a_2t^2)^{2^{m+1}}={d^2(a_1t+a_2t^2)^2}+ {d^2}{w(a_1t+a_2t^2)}.  
\end{equation*}
Using (\ref{e38}), we have
\begin{equation*}\label{eb2}
(a_1^2(t^2+t)+a_2^2t )={d^2(a_1^2t^2+a_2^2(t^2+t))}+ {d^2}{w(a_1t+a_2t^2)}.  
\end{equation*}
The above equation is equivalent to the following two formulars
\begin{equation}\label{eb3}
a_1^2 +d^2(a_1^2+a_2^2)+d^2wa_2=0, 
\end{equation}
and
\begin{equation}\label{eb4}
a_1^2+a_2^2+ d^2a_2^2+d^2wa_1=0.
\end{equation}
Then $(\ref{eb3})*a_1+(\ref{eb4})*a_2$ is equal to
\begin{equation*}\label{ebb1}
a_1^3 +d^2(a_1^3+a_1a_2^2+a_2^3)+a_1^2a_2+a_2^3 =0. 
\end{equation*}
By equation (\ref{e48}), the above equation can be transformed into
\begin{equation*}\label{ebb2}
a_1^3 +d^2 +a_1^2a_2+a_2^3  =0.
\end{equation*}
Multiply $a_1$ for (\ref{e54}) 
\begin{equation*}\label{ebb3}
d_1a_1=a_1^3+a_1a_2^2+a_1^2a_2.
\end{equation*}
Addding the above two equations, we have
\begin{equation*}\label{ebb4}
a_1^3+d_1a_1+d^2+(a_1a_2^2+ a_1^3 +  a_2^3)=0.
\end{equation*}
By equation (\ref{e48}), the above equation ie equivalent to
\begin{equation}\label{ebb5}
a_1^3+d_1a_1+d^2+ 1=0.
\end{equation}
That is both $a_1$ and $a_2$ satisfy equation (\ref{ebb5}).
Now, the result follows from Lemma \ref{l01}.

  Necessities: 

As above, we assume that $d\not=1$. Now, let $c\in \mathbb{F}_{2^{m}}^*$ satisfies $p_m({{c^2+1}\over {  1+c+c^2}})=0$. For equations (\ref{e22}) and (\ref{e23}), we need to find $a_1$ and $ a_2$ not both zeroes, such that (\ref{eb3}) and (\ref{eb4}) are satisfied.

 Set $d={1\over c}, d_1=(1+d^2+d^4)^{1\over 3}, w={d_1^2\over d^2}$. Then according to Lemma \ref{l01}, 
\begin{equation}\label{eb5}
a^3+d_1a+d^2+1=0
\end{equation}
has three distinct roots in $ \mathbb{F}_{2^{m}}^*$. Let $a_1$ be one of the roots. Set 
\begin{equation*}\label{eb6}
a_2={{a_1^3+1}\over {d_1+a_1^2}}.
\end{equation*}
Note that if $d_1=a_1^2$, then (\ref{eb5}) implies that $d=1$ contradiction. Substituting $a_2$ into the left hand side of (\ref{eb3}), we have
\[
{{a_1^6 + a_1^5d^2w + a_1^3d^2d_1w + a_1^2d^2d_1^2 + a_1^2d^2w + a_1^2d_1^2 +
d^2*d_1w + d^2 }\over {(a_1^2+d_1)^2}}.
\]
Combining with the definition of $w$, it becomes
\begin{equation}\label{eb8}
{{a_1^6 + a_1^5d_1^2 + a_1^3d_1^3 + a_1^2d^2d_1^2 + d^2 + d_1^3 }\over {(a_1^2+d_1)^2}}.
\end{equation}
Now, since $a_1$ satisfies (\ref{eb5}), we have 
\begin{equation*}\label{eb9}
a_1^3=d_1a_1+d^2+1.
\end{equation*}
Substituting into (\ref{eb8}),  we find that it is zero which is   the right side of (\ref{eb3}). Note that since $m$ is odd, we have $1+d+d^2\not=0$, that is $d_1\not=0$. We can also find that $a_1, a_2$ satisfy (\ref{eb4}).

As for the case $m=3k+2$, corresponding to (\ref{eb3}) and (\ref{eb4}), we need to consider the following two equatioins
\begin{equation*} 
a_2^2 +d^2(a_1^2+a_2^2)+d^2wa_2 =0,
\end{equation*}
and
\begin{equation*} 
a_1^2+  d^2a_2^2+d^2wa_1=0.
\end{equation*}
Instead of equaiton (\ref{ebb5}), $a_1$ and $a_2$ satisfy  
\begin{equation*} 
a^3+d_1a+d^2=0.
\end{equation*}
If   $c=d=1$, then $d_1=1$ and the above equation has no solutions in $\mathbb{F}_{2^m}$.

Finally, for $c=d=1$, it is not difficult to find that  the number of terms in $p_k(x)$ is of the following form
\[
odd, \ odd, even, odd, odd, even ,odd, odd,even, \cdots,
\]
that is if $k$ is a multiple of $3$, it has  even number of terms, otherwise it has odd number of terms. Thus for $m=3k+2$, $p_m({{1}\over {  1+c+c^2}})=p_{3k+2}(1)=1\not=0$, $f(x)$ is a PP. But for $m=3k+1$, $p_m({{c^2+1}\over {  1+c+c^2}})=p_m(0)=0$, $f(x)$ is not a PP. Thus the $c=1$ situation is consistent with the general case.
\end{IEEEproof}

Combining with \cite[Proposition 3]{ZYY}, we have the following result.
\begin{corollary}\label{c01}
Let $3\nmid m$ be a positive odd integer, and  $\delta \in \mathbb{F}_{2^{3m}}, c\in \mathbb{F}_{2^{m}}^*$. The polynomial
\[
g(x)=x^{2^{2m}+2^m}+x^{2^{2m}+1}+c^2x^2
\]
\begin{enumerate}
\renewcommand{\labelenumi}{$($\mbox{\roman{enumi}}$)$}
\item
  is a PP of $\mathbb{F}_{2^{3m}}$ if and only  $p_m({{c^2+1}\over {  1+c+c^2}})\not=0$ when $m=3k+1$;
\item
 is a PP of $\mathbb{F}_{2^{3m}}$ if and only if  $p_m({{1}\over {  1+c+c^2}}) \not=0$ when $m=3k+2$.
\end{enumerate}
\end{corollary}

\begin{example}
Let $m=7$. Using Magma, it can be found that
\[ 
p_7(x)=x^{22} + x^{21} + x^{19} + x^{18} + x^{17} + x^{11} + x^{10} + x^9 + x^6 + x^5 + x^3 + x^2 + x.
\]
There are $84$ elements $c\in \mathbb{F}_{2^7}^*$ satisfying $p_7({{c^2+1}\over {  1+c+c^2}}) \not=0$, for all such $c$
\begin{equation}\label{ex1}
g=x^{16385}+x^{16512}+c^2x^2
\end{equation}
is a PP over $\mathbb{F}_{2^{21}}$.  For the remaining $43$ elements, (\ref{ex1}) is not a PP.

\end{example}

\begin{example}
Let $m=5$. Using Magma, it can be found that  
\[ 
p_5(x)=x^6 + x^5 + x^3 + x^2 + x.
\]
There are $21$ elements $c\in \mathbb{F}_{2^5}^*$ satisfying $p_({{1}\over {  1+c+c^2}}) \not=0$, for all such $c$
\begin{equation}\label{ex2}
g=x^{16385}+x^{16512}+c^2x^2
\end{equation}
is a PP over $\mathbb{F}_{2^{15}}$.
For the remaining $10$ elements, (\ref{ex2}) is not a PP.

\end{example}

\section{Conclusion}



Permutation polynomials over finite fields are interesting for their simple algebraic forms, and they have many applications in areas of mathematics and engineering. They can be used to construct linear codes and cyclic codes, they can be employed in bent and semi-bent functions.  In this paper, we find the necessties and sufficiencies for a class of polynomials to be PPs. 


 \section*{Acknowledgment}

 The author would like to thank the anonymous referees for helpful suggestions and comments.


\end{document}